\documentclass{ws-p9-75x6-50}

\renewcommand{\rightnote}{%
\textit{Hadron 13 : Precision Tests of Electroweak Physics}}

\newcommand{\AQ}[1]{\Pi_{\rm #1}(q^2)}
\newcommand{\AZERO}[1]{\Pi_{\rm #1}(0)}

\newcommand{\APZERO}[1]{\Pi'_{\rm #1}(0)}
\newcommand{\SM}[1]{\left[ #1 \right]_\mathrm{SM}}
\newcommand{\seff}{\sin^2\theta_\mathrm{eff}^\mathrm{lept}}
\newcommand{\Gaml}{\Gamma_{\ell^+\ell^-}}
\newcommand{\QWCS}{Q_W({}^{133}_{\phantom{1}55}{\rm Cs})}
\newcommand{\QWTL}{Q_W({}^{205}_{\phantom{1}81}{\rm Tl})}

\newcommand{\dfrac}[2]{\frac{\displaystyle #1}{\displaystyle #2}}

\newcommand{\hepex}[1]{hep--ex/#1}
\begin{document}

\setcounter{page}{0}
\pagestyle{empty}

\begin{flushright}
VPI--IPPAP--99--03\\
hep--ph/yymmddd\\
April 1999
\end{flushright}

\bigskip
\bigskip
\bigskip
\bigskip

\begin{center}

\textbf{\Large Precision Tests of Electroweak Physics}

\bigskip
\bigskip
\bigskip

\textsc{\large Tatsu~TAKEUCHI}\footnote{%
electronic address: takeuchi@vt.edu},
\textsc{\large Will~LOINAZ}\footnote{%
electronic address: loinaz@alumni.princeton.edu}\\
\medskip
\textit{Institute for Particle Physics and Astrophysics\\
Physics Department, Virginia Tech, Blacksburg, VA 24061, USA}\\
\bigskip
and\\
\bigskip
\textsc{\large Aaron~K.~Grant}\footnote{%
electronic address: grant@gauss.harvard.edu}\\
\medskip
\textit{Physics Department, Harvard University, Cambridge, MA 02138, USA}\\

\bigskip
\bigskip
\bigskip

\begin{quote}
We review the current status of precision electroweak measurements and the
constraints they impose on new physics.  We perform a model
independent analysis using the $STU$--formalism of Ref.~1, and then
discuss how the $Z$--pole data from LEP and SLD can be used
to constrain models that are not covered within that framework.
\end{quote}

\bigskip
\bigskip
\bigskip

\textit{Talk presented by Tatsu Takeuchi at\\
Hadron Collider Physics 13, Mumbai, India, January 14--20, 1999}

\end{center}

\vfill

\begin{flushleft}
VPI--IPPAP--99--03\\
hep--ph/yymmddd\\
April 1999
\end{flushleft}

\setcounter{footnote}{1}
\newpage
\pagestyle{plain}

\title{Precision Tests of Electroweak Physics}

\author{Tatsu Takeuchi\footnote{Presenting author.} and Will Loinaz}

\address{Institute for Particle Physics and Astrophysics\\
Physics Department, Virginia Tech, Blacksburg, VA 24061, USA}

\author{Aaron K. Grant}

\address{Physics Department, Harvard University, 
Cambridge, MA 02138, USA}  

\maketitle

\abstracts{We review the current status of precision
electroweak measurements and the constraints they impose
on new physics.  We perform a model independent analysis using
the $STU$--formalism of Ref.~1, and
then discuss how the $Z$--pole data from LEP and SLD can be 
used to constrain models that are not covered within that
framework.
}

\section{Introduction}

There are two reasons for studying the data from precision 
electroweak measurements.
First, we would like to know how well the
Standard Model (SM) agrees with the experimental data:
any significant deviation between the two
would be a signal of new physics beyond the SM.
Also, since the SM predictions depend on the mass of
the still unobserved Higgs boson, the value
of $m_H$ which best reproduces the experimental data
serves as its prediction.
Second, we would like to know 
if any of the proposed theories of new physics beyond the SM
is viable, i.e. does not
make theoretical predictions inconsistent with
the experimental data (or perhaps makes the agreement
even better than with the SM).

The way to kill these two birds with one stone is to
constrain the sizes of radiative corrections coming
from new physics using the precision electroweak data.
If the data prefers a non--zero value for the size
of these extra corrections for some choice of $m_H$, 
it would signal that:
(1) the agreement between the SM and data is not
perfect for that particular choice of Higgs mass, and
(2) any new physics which predicts such corrections would 
make the agreement between theory and experiment better.

\section{Constraints on Oblique Electroweak Corrections}

In order to constrain the size of radiative corrections coming
from new physics, we must first make some assumptions about the new
physics giving rise to them (since it is impossible to consider
all possible corrections from all possible new physics all at once) 
and express them in terms of a few model independent parameters.

\begin{figure}[t]
\setlength{\unitlength}{1mm}
\begin{picture}(130,55)(-65,-10)
\put(-40,0){\line(1,0){80}}
\put(0,-10){\vector(0,1){50}} 
\put(-5,42){Energy}
\put(-32,-5){Charged Channel}
\put(5,-5){Neutral Channel}
\put(-22,4){$G_\mu$}
\put(18.8,4.2){$\alpha$}
\put(-22,28){$M_W$}
\put(17.3,33.2){$M_Z$}
\put(0,5){\vector(-1,0){15}}
\put(0,5){\vector(1,0){15}}
\put(20,20){\vector(0,-1){10}}
\put(20,19){\vector(0,1){10}}
\put(-20,20){\vector(0,-1){10}}
\put(-20,20){\vector(0,1){5}}
\put(3,6.5){$T$}
\put(21.5,19){$S$}
\put(-30.5,15){$S+U$}
\put(-20,5){\circle{7}}
\put(20,5){\circle{7}}
\put(20,34){\circle{7}}
\end{picture}
\caption{The $S$, $T$, and $U$ variables quantify the extra
vacuum polarization corrections due to new physics when relating
physics in different channels or physics at different energy scales.
Low energy and neutral current observables depend only on $S$ and
$T$ because $\alpha$, $G_\mu$, and $M_Z$ are used as inputs
to make the SM predictions. \label{FIG1}}
\end{figure}
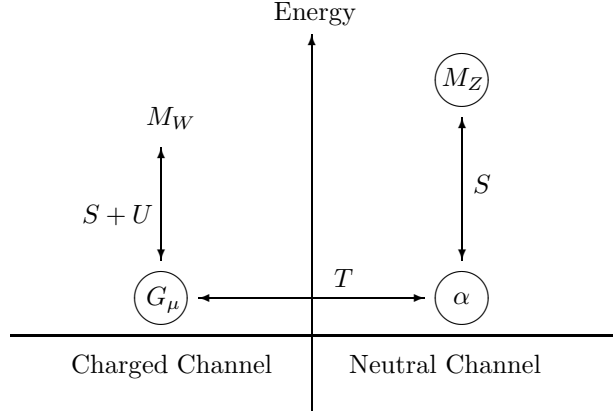

The simplest assumptions that would sufficiently constrain the
number of corrections one must consider while still encompassing 
a large class of models are the following:
\begin{enumerate}
\item{The electroweak gauge group is the standard
      $SU(2)_L \times U(1)_Y$.  The only electroweak gauge
      bosons are the photon, the $W^\pm$, and the $Z$.}
\item{The couplings of new physics to light fermions are highly
      suppressed.  Since all precision electroweak measurements
      only involve four fermion processes with light external
      fermions, this means that vertex and box corrections from 
      new physics can be neglected. 
      Only vacuum polarization
      (i.e. oblique) corrections need to be considered.  
      }
\item{The mass scale of new physics is large compared to the $W$
      and $Z$ masses.}
\end{enumerate}
The first two assumptions let us focus our attention on
just four vacuum polarization corrections: 
$\AQ{WW}$, $\AQ{ZZ}$, $\AQ{Z\gamma}$, and $\AQ{\gamma\gamma}$.
Here, $\AQ{XY}$ is the transverse part of the vacuum polarization
function between gauge bosons $X$ and $Y$.
The third assumption lets us expand these vacuum polarization functions
around $q^2=0$ and neglect the higher order terms since they
are suppressed by powers of $q^2/M_\mathrm{new}^2$ where $q^2 \le M_Z^2$
for the processes under consideration:
\begin{eqnarray*}
\AQ{WW} & = & \AZERO{WW} + q^2 \APZERO{WW} + \cdots \cr
\AQ{ZZ} & = & \AZERO{ZZ} + q^2 \APZERO{ZZ} + \cdots \cr
\AQ{Z\gamma} & = & q^2 \APZERO{Z\gamma} + \cdots \cr
\AQ{\gamma\gamma} & = & q^2 \APZERO{\gamma\gamma} + \cdots
\end{eqnarray*}
Therefore, we can express the radiative corrections from new
physics in terms of just six parameters:
$\AZERO{WW}$, $\APZERO{WW}$, $\AZERO{ZZ}$, $\APZERO{ZZ}$,
$\APZERO{Z\gamma}$, and $\APZERO{\gamma\gamma}$.
Of these six, three will be absorbed into the renormalization of
the three input parameters $\alpha$, $G_\mu$ and $M_Z$,
leaving us with three observables parameters which can be taken 
to be: \cite{PESKIN:90}
\begin{eqnarray*}
\alpha S & = & 4s^2 c^2
               \left[ \APZERO{ZZ}
                      -\frac{c^2-s^2}{sc}\APZERO{Z\gamma}
                      -\APZERO{\gamma\gamma}
               \right]\,,  \nonumber \\
\alpha T & = & \frac{\AZERO{WW}}{M_W^2} - \frac{\AZERO{ZZ}}{M_Z^2}\,, \\
\alpha U & = & 4s^2
               \left[ \APZERO{WW} - c^2\APZERO{ZZ}
                      - 2sc\APZERO{Z\gamma} - s^2\APZERO{\gamma\gamma} 
               \right]\,. \nonumber
               \phantom{\frac{e^2}{s^2}}   
\end{eqnarray*}
Here, $\alpha$ is the fine structure
constant and $s$ and $c$ are the sine and
cosine of the weak mixing angle.  
Only the contribution of new physics to these functions are to be included.
The parameters $T$ and $U$ are defined so that
they vanish if new physics does not break custodial $SU(2)$
symmetry. See Ref.~2 for a discussion on the symmetry
properties of $S$.

\begin{table}[t]
\begin{center}
\begin{tabular}{|l|c|c|c|}
\hline
Observable    & SM prediction & Measured Value & Reference\\
\hline\hline
\underline{$\nu_\mu e$ and $\bar{\nu}_\mu e$ scattering} & & & \\ 
$g_V^{\nu e}$ & $-0.0365$     & $-0.041  \pm 0.015$  & \cite{PDG:98} \\
$g_A^{\nu e}$ & $-0.5065$     & $-0.507  \pm 0.014$  & \cite{PDG:98} \\
\hline
\underline{Atomic Parity Violation} & & & \\
$\QWCS$       & $-73.19$      & $-72.41  \pm 0.84$   & \cite{PDG:98} \\
$\QWTL$       & $-116.8$      & $-114.8  \pm 3.6$    & \cite{PDG:98} \\
\hline
\underline{$\nu_\mu N$ and $\bar{\nu}_\mu N$ DIS} & & & \\
$g_L^2$       & $0.3031$      & $0.3009  \pm 0.0028$ & \cite{PDG:98} \\
$g_R^2$       & $0.0304$      & $0.0328  \pm 0.0030$ & \cite{PDG:98} \\
NuTeV         & $0.2289$      & $0.2277  \pm 0.0022$ & \cite{NUTEV:98} \\
\hline
\underline{LEP/SLD} & & & \\
$\Gaml$       & $0.08392$ GeV & $0.08390 \pm 0.00010$ GeV & \cite{LEP:98} \\
$\seff$ (LEP) & $0.23200$     & $0.23153 \pm 0.00034$     & \cite{LEP:98} \\
$\seff$ (SLD) & $0.23200$     & $0.23109 \pm 0.00029$     & \cite{LEP:98} \\
\hline
\underline{$W$ mass} & & & \\
$M_W$ ($p\bar{p}$ + LEP2)
              & $80.315$ GeV  & $80.39 \pm 0.06$ GeV    & 
\cite{LEP:98} \\
\hline
\end{tabular}
\caption{
The data used for the oblique correction analysis.
The value of $\seff$ for LEP is from leptonic asymmetries only.
The SM predictions for the $W$ mass and the LEP/SLD observables 
were obtained using the program ZFITTER 4.9 \protect\cite{ZFITTER:92}.
The predictions for the low energy observables were 
calculated from the formulae given in Ref.~7.  
The parameter choice for the reference SM was 
$M_Z = 91.1867$ GeV\protect\cite{LEP:98}, 
$m_t = 173.9$ GeV\protect\cite{TOPMASS:98}, 
$m_H = 300$ GeV, $\alpha^{-1}(M_Z) = 128.9$\protect\cite{ALEMANY:98},
and $\alpha_s(M_Z) = 0.120$. 
\label{DATA1}}
\end{center}
\end{table}

The fact that three parameters are necessary to describe
the effects of new physics can also be understood as follows:
Since vacuum polarizations modify the gauge boson propagators,
their presence can be seen when comparing the exchange of
different electroweak gauge bosons, or when comparing the exchange
of the same boson at different energy scales.
In order to make predictions based on the three input parameters
$\alpha$, $G_\mu$, and $M_Z$, which are
neutral current--low energy, charged current--low energy, and
neutral current--high energy observables, respectively,
one must compare the theory in the charged ($W$ exchange)
and the neutral ($Z$ and photon exchange) channels,
as well as in the same channel at different energy scales
as shown by arrows in Fig.~\ref{FIG1}.  New physics effects
will manifest themselves in each of the three arrows
shown in Fig.~\ref{FIG1}.    The parameter $S$ quantifies the
extra correction from new physics one must include when comparing
neutral current processes at different energy scales while the
parameter $T$ quantifies the extra corrections that must be
included when comparing charged and neutral current processes
at low energy.   A third parameter, in this case $S+U$, is 
necessary to quantify the correction due to
new physics when comparing charged channel processes at
different energy scales.

This discussion also shows that all low energy and
neutral current observables will receive extra corrections
from only $S$ and $T$.   Of all the precision electroweak
measurements, the only quantity which receives a correction from $U$ 
is the $W$ mass.
Therefore, the majority of the precision data can be used to
constrain just two parameters $S$ and $T$, which in turn
can be calculated for each model of new physics beyond the SM.

The details of how to calculate the corrections to
various observables from  $S$, $T$, and $U$
can be found elsewhere\cite{PESKIN:90}.  One obtains expressions
such as
\[
M_W/\SM{M_W}
= 1 + \frac{\alpha}{2(c^2-s^2)}
      \left[ -\frac{1}{2}S + c^2 T + \frac{c^2-s^2}{4s^2} U
      \right],
\]
where $\SM{M_W}$ is the SM prediction of $M_W$.
These expressions can be compared directly with the experimental data
to place constraints on $S$, $T$, and $U$.

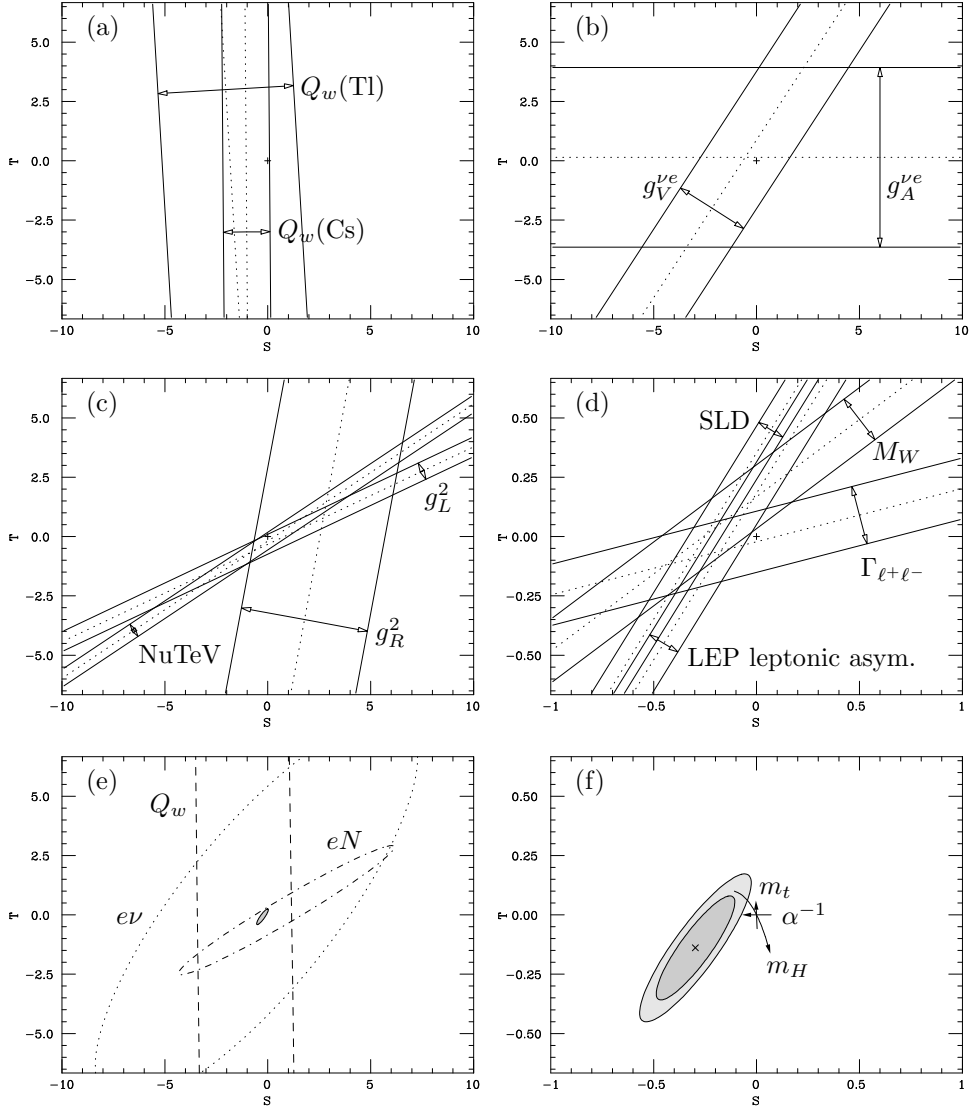
\begin{figure}[t]
\setlength{\unitlength}{1mm}

\begin{picture}(130,100)(0,0)
\put(0,50){
\epsfxsize=2.5in 
\epsfbox{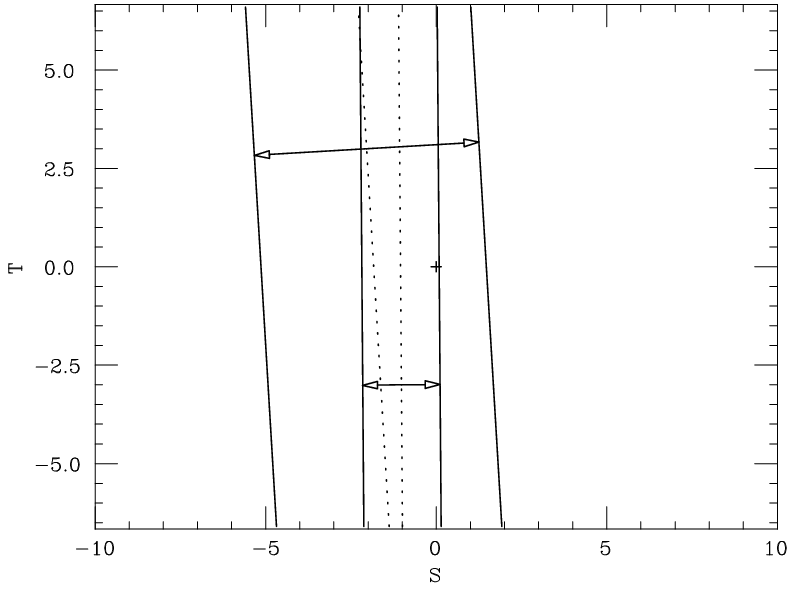}
}
\put(40.5,85){$Q_w(\mathrm{Tl})$}
\put(37.5,66){$Q_w(\mathrm{Cs})$}
\put(65,50){
\epsfxsize=2.5in 
\epsfbox{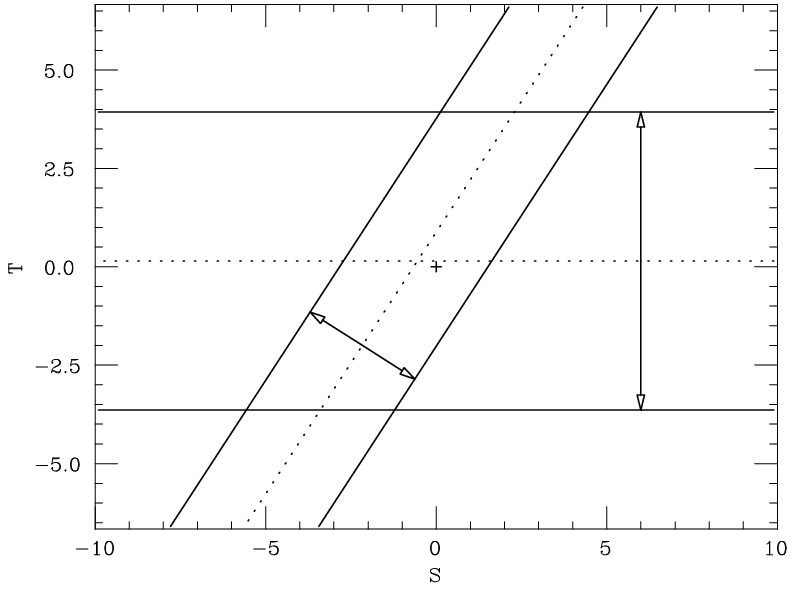}
}
\put(86,72){$g_V^{\nu e}$}
\put(118.5,72){$g_A^{\nu e}$}
\put(0,0){
\epsfxsize=2.5in 
\epsfbox{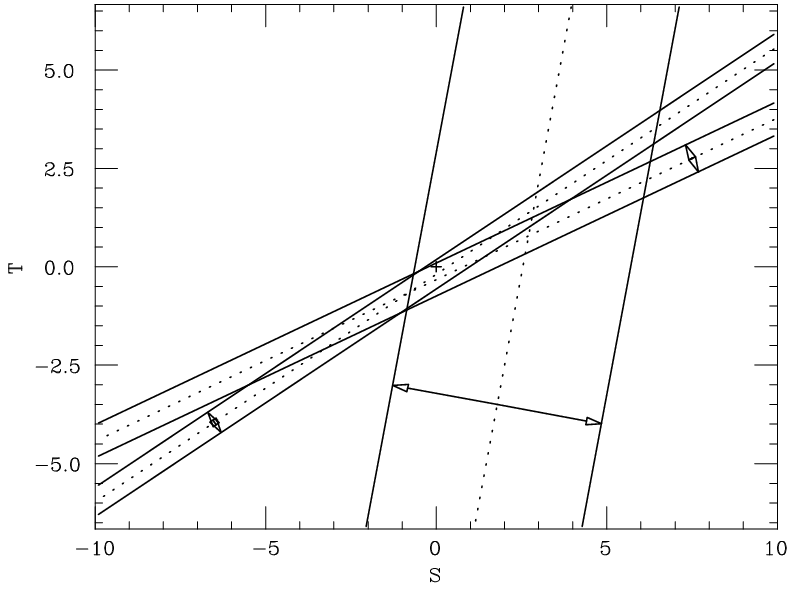}
}
\put(57,30.5){$g_L^2$}
\put(50.5,12.5){$g_R^2$}
\put(19,9.5){NuTeV}
\put(65,0){ 
\epsfxsize=2.5in 
\epsfbox{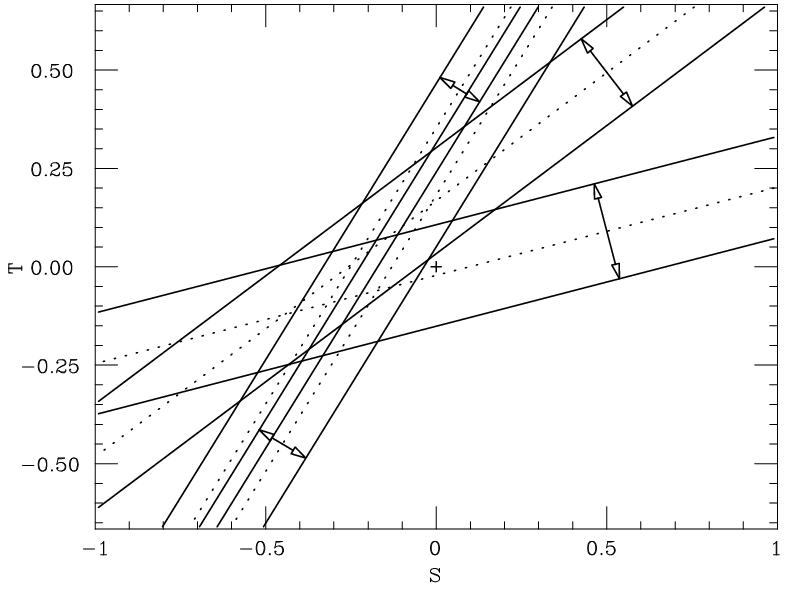}
}
\put(92,9){LEP leptonic asym.}
\put(93.5,40.5){SLD}
\put(116.5,36.5){$M_W$}
\put(115,21){$\Gamma_{\ell^+\ell^-}$}

\put(12,93){(a)}
\put(77,93){(b)}
\put(12,43){(c)}
\put(77,43){(d)}
\end{picture}

\begin{picture}(130,50)(0,0)
\put(0,0){
\epsfxsize=2.5in
\epsfbox{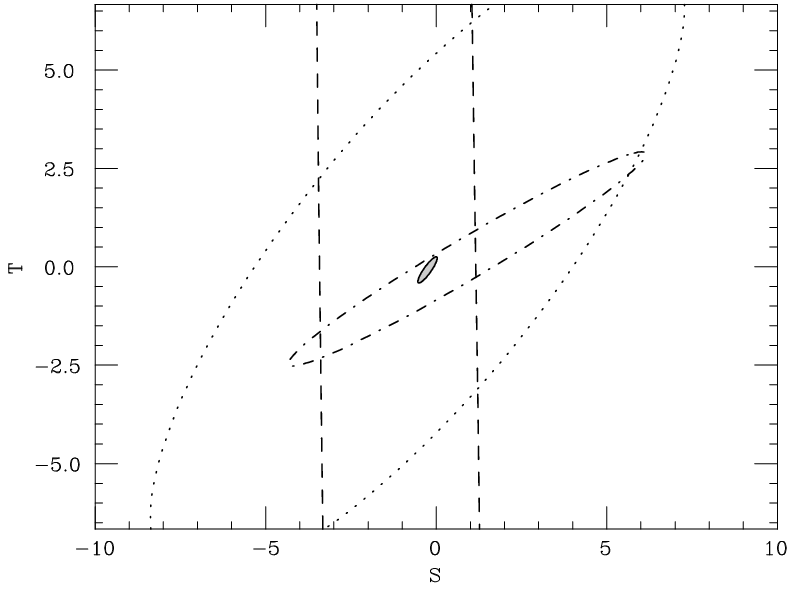}
}
\put(16,25){$e\nu$}
\put(44,35){$eN$}
\put(20.5,40){$Q_w$}
\put(65,0){
\epsfxsize=2.5in
\epsfbox{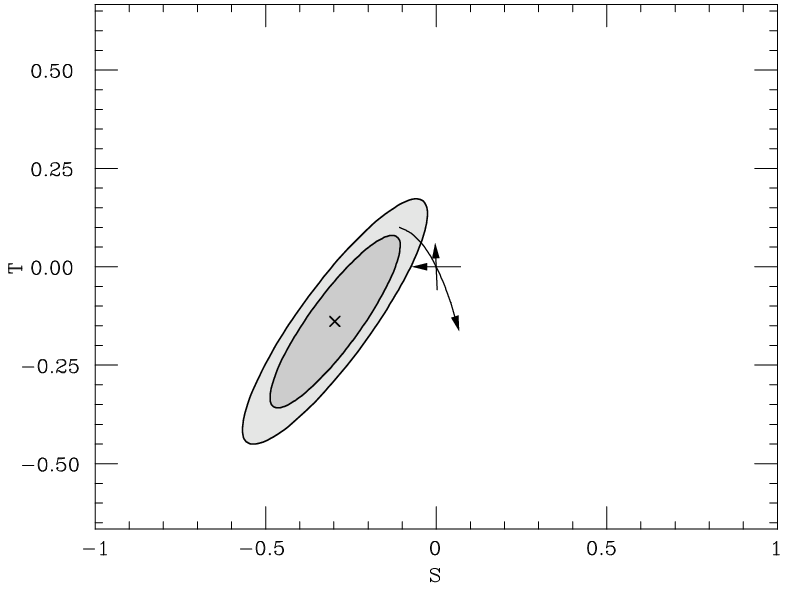}
}
\put(101.5,29){$m_t$}
\put(102.5,18.5){$m_H$}
\put(104.5,25){$\alpha^{-1}$}

\put(12,43){(e)}
\put(77,43){(f)}
\end{picture}

\caption{The 1--$\sigma$ limits on $S$ and $T$ from 
(a) atomic parity violation,
(b) $e\nu_\mu$ and $e\bar{\nu}_\mu$ scattering,
(c) $\nu_\mu$ and $\bar{\nu}_\mu$ DIS, and
(d) LEP/SLD + $M_W$.
(e) The 90\% confidence contours for the four classes
of experiments. The LEP/SLD + $M_W$ contour is the small
shaded area in the middle.
(f) The 68\% and 90\% confidence limits on $S$ and $T$, 
all experiments combined.  The arrows indicate how the SM point will
move relative to the contours for $m_t = 173.9 \pm 5$ GeV, 
$m_H = 300^{+700}_{-210}$ GeV, $\alpha^{-1}(M_Z) = 128.9 \pm 0.1$.
\label{FIG2}}
\end{figure}

In Table~\ref{DATA1} we list the data we used in our analysis.
The definitions of the parameters $g_{V/A}^{e\nu}$ for
$e\nu_\mu$ and $e\bar{\nu}_\mu$ scattering, the weak charge $Q_w$ for
atomic parity violation, and $g_{L/R}^2$ for $\nu_\mu$ and $\bar{\nu}_\mu$
deep inelastic scattering (DIS) can be found in the
Review of Particle Physics\cite{PDG:98} from which the data were
taken.
The quantity measured by the NuTeV collaboration\cite{NUTEV:98}
is a linear combination of $g_L^2$ and $g_R^2$ for which the 
uncertainty due to the charm threshold cancels.
The rest of the data is from Ref.~5.

Comparing the experimental data to SM predictions with
$m_t = 173.9$~GeV\cite{TOPMASS:98},
$m_H = 300$~GeV, and
$\alpha^{-1}(M_Z) = 128.9$\cite{ALEMANY:98},
we obtain the constraints shown in Fig.~\ref{FIG2}.
Note that Figs.~\ref{FIG2}d and \ref{FIG2}f are 
drawn at a different scale from the other four.
Fig.~\ref{FIG2}e shows the 90\% confidence
limits on $S$ and $T$ due to the four classes of experiments separately, 
and Fig.~\ref{FIG2}f shows the 68\% and 90\% confidence limits
from all experiments combined.
As is evident from Fig.~\ref{FIG2}, the LEP/SLD measurements
provide the tightest constraints on $S$ and $T$.
All of the other observables combined have little effect on the final
result, which is
\begin{eqnarray*}
S & = & -0.30 \pm 0.13, \cr
T & = & -0.14 \pm 0.15, \cr
U & = & \phantom{-}0.15 \pm 0.21. 
\end{eqnarray*}
These limits of course depend on the values of $m_t$, $m_H$, and
$\alpha^{-1}(M_Z)$ used as input to calculate the SM predictions.
The dependence of the limits on these input parameters is shown
by arrows in Fig.~\ref{FIG2}f.
We can see that the current data favor
either a small value of the Higgs mass or a larger value of
$\alpha^{-1}(M_Z)$.

\section{Limits on Topcolor Assisted Technicolor}

\begin{table}[t]
\begin{center}
\begin{tabular}{|c||c|c|c|c|c|}
\hline
           & $SU(3)_s$ & $SU(3)_w$ & $U(1)_s$ & $U(1)_w$ & $SU(2)_L$ \\
\hline\hline
$(t,b)_L$  &  3  &  1  & $\dfrac{1}{3}$  &   0   &   2   \\
\hline
$(t,b)_R$  &  3  &  1  & $\left(\dfrac{4}{3},-\dfrac{2}{3}\right)$
                                        &   0   &   1   \\
\hline
$(\nu_\tau,\tau^-)_L$
           &  1  &  1  & $-1$     &   0   &   2   \\
\hline
$\tau^-_R$ &  1  &  1  & $-2$     &   0   &   1   \\
\hline
$(c,s)_L$, 
$(u,d)_L$  &  1  &  3  &  0  & $\dfrac{1}{3}$   &   2   \\
\hline
$(c,s)_R$, 
$(u,d)_R$  &  1  &  3  &  0  & $\left(\dfrac{4}{3},-\dfrac{2}{3}\right)$
                                                &   1   \\
\hline
$(\nu_\mu,\mu^-)_L$, $(\nu_e,e^-)_L$
           &  1  &  1  &  0  & $-1$  &   2   \\
\hline
$\mu^-_R$, 
$e^-_R$    &  1  &  1  &  0  & $-2$  &   1   \\
\hline
\end{tabular}
\end{center}
\caption{Charge assignments of the ordinary fermions
in topcolor assisted technicolor.\label{CHARGES}}
\end{table}

The limits on $S$ and $T$ are useful in constraining new
physics models which satisfy the three initial assumptions
but are less useful for other theories.
As an example of such a theory, let us consider
topcolor assisted technicolor\cite{HILL:95}
with the gauge group
\[
SU(3)_s \times SU(3)_w \times U(1)_s \times U(1)_w \times SU(2)_L
\]
The coupling constants for the two $SU(3)$'s and the two $U(1)$'s
are assumed to satisfy $g_{3s}\gg g_{3w}$ and $g_{1s}\gg g_{1w}$.
The charges of the ordinary fermions under these groups are shown
in Table~\ref{CHARGES}.

At a scale of about one TeV, it is assumed that technicolor
breaks the two $SU(3)$'s and the two $U(1)$'s to their diagonal
subgroups:
\[
SU(3)_s \times SU(3)_w \rightarrow SU(3)_c,\qquad
U(1)_s \times U(1)_w   \rightarrow U(1)_Y,
\]
which are identified with the SM color and hypercharge groups.   
Because of the assumption $g_{3s}\gg g_{3w}$ and $g_{1s}\gg g_{1w}$,
the broken $SU(3)$ gauge bosons (the \textit{colorons})
and the broken $U(1)$ gauge boson (the $Z'$)
couple strongly to the third generation fermions
but only weakly to the first and second generation fermions.
Coloron exchange is attractive in both the $t\bar{t}$ and
$b\bar{b}$ channels, while $Z'$ exchange is attractive
for $t\bar{t}$ but repulsive for $b\bar{b}$.
The combined strength of the coloron and $Z'$ interactions
is assumed to be strong enough to condense the top but
not so strong as to condense the bottom.
As a result, only the top quark becomes heavy.

It is easy to see that this model does not fit into the
$STU$ framework since (1) it has an extra electroweak gauge
boson, the $Z'$, and (2) coloron and $Z'$ exchange
can lead to large vertex corrections for the third
generation fermions ($b$, $\tau$, and $\nu_\tau$).
How would we place constraints on such a model?

A naive extension of the $STU$ formalism to include
the $Z'$ vacuum polarization functions turns out to be
too complicated to be illuminating.
A much better way is to concentrate our attention
on the vertex corrections at the $Z$ mass scale,
where we have a wealth of data from LEP and SLD.
Recall from our previous discussion that
$S$ and $T$ are relevant only when comparing
processes at different energy scales or in different channels.
If we only look at the LEP/SLD data, which come
from neutral current processes at the $Z$ mass scale,
we can make our analysis completely blind to the
vacuum polarization corrections and obtain limits on
the vertex corrections only.
Another way to see this is to notice that most of
the observables at LEP/SLD are asymmetries and
branching fractions which are just
\textit{ratios of coupling constants} at the $Z$ mass scale.
The SM predictions for these observables can be fixed by using
only one of them as input to predict all the others,
and any deviations must come from 
the vertex corrections due to new physics.\footnote{%
We have used a similar technique in Ref.~11.}

Since the details of our analysis has been presented
elsewhere\cite{LOINAZ:98}, we give only an outline here.
In the topcolor assisted technicolor model considered above,
vertex corrections come in two classes: (1) gauge boson mixing terms,
and (2) proper vertex corrections.
Gauge boson mixing modifies the current to which the $Z$ couples to
from $J_Z^0 = J_{I_3} - s^2 J_Q$ to
\[
J_Z = J_{I_3} - (s^2 + \delta s^2) J_Q + \epsilon J_{1s}.
\]
The parameters $\delta s^2$ and $\epsilon$ quantify the amount
of $Z$--photon and $Z$--$Z'$ mixing, respectively.
The relevant proper vertex corrections are the coloron and
$Z'$ corrections to the third generation fermion vertices\cite{HILL:95A}, 
the sizes of which we parametrize by 
\[
\kappa_i = 
\frac{g_{is}^2}{4\pi}
\left( \frac{g_{is}^2}{g_{is}^2+g_{iw}^2} \right),
\qquad (i=1,3),
\]
and a correction to the left--handed coupling of the $b$ to
the $Z$ from the top--pion loop\cite{BURDMAN:97}
which we denote $\Delta$.

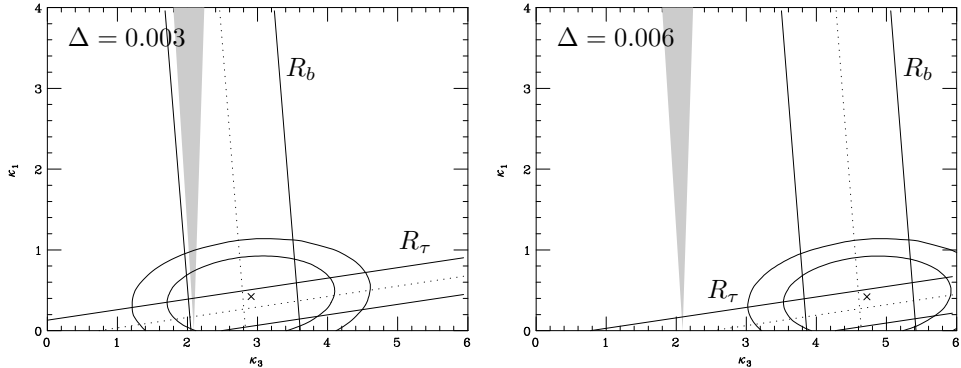
\begin{figure}[t]
\setlength{\unitlength}{1mm}
\begin{picture}(130,50)(0,0)
\put(0,0){
\epsfxsize=2.5in
\epsfbox{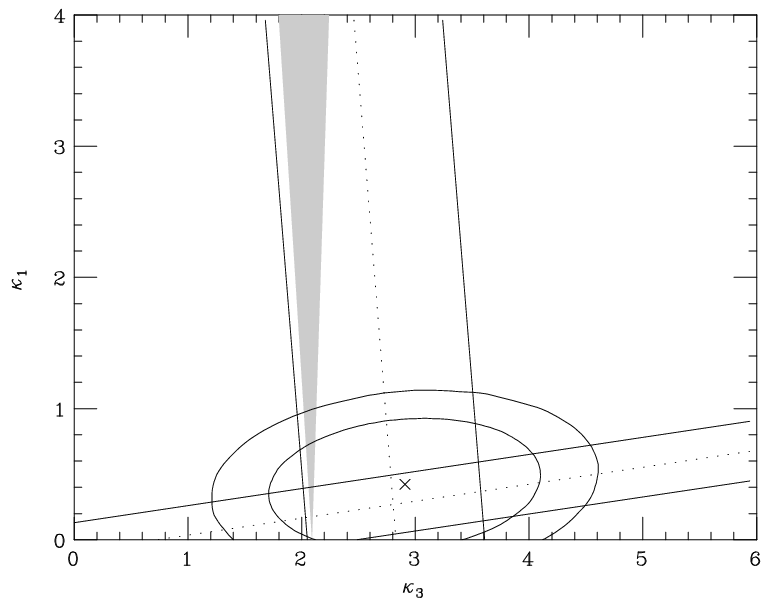}
}
\put(39,40){$R_b$}
\put(54,17){$R_\tau$}
\put(10,44){$\Delta=0.003$}
\put(65,0){
\epsfxsize=2.5in
\epsfbox{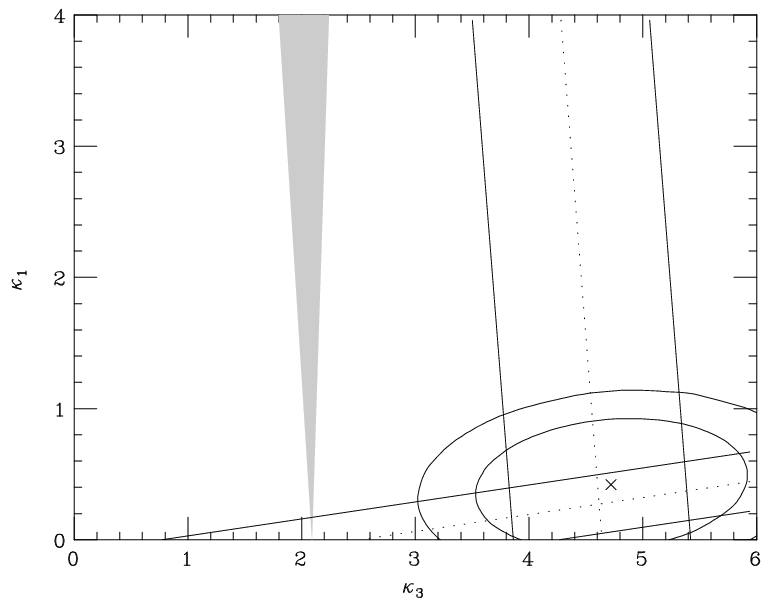}
}
\put(121,40){$R_b$}
\put(95,10.5){$R_\tau$}
\put(75,44){$\Delta=0.006$}
\end{picture}
\caption{Limits on $\kappa_1$ and $\kappa_3$ from 
LEP/SLD observables. $\Delta$ parametrizes the
size of the top--pion correction to the $Zb\bar{b}$ vertex.
The shaded region is where the combined coloron and $Z'$
interaction can condense the top without condensing the bottom
or the $\tau$.
\label{FIG3}}
\end{figure}

The corrections to various LEP/SLD obserables from $\delta s^2$,
$\epsilon$, $\kappa_1$, $\kappa_3$, and $\Delta$ were calculated and
compared to the experimental data.
In performing the fit, we kept the value of $\Delta$ fixed
and let the four other parameters and the QCD coupling constant
$\alpha_s(M_Z)$ float.
In Fig.~\ref{FIG3}, we show only the results in
the $\kappa_1$--$\kappa_3$ plane for two choices of the value of
$\Delta$: $0.003$ and $0.006$.
These correspond to top--pion masses of $m_+ = 1000$~GeV and
$m_+ = 600$~GeV, respectively.
$\kappa_1$ and $\kappa_3$ must fall into the shaded region 
in order for the coloron and $Z'$ interactions to condense the
top while not condensing the bottom.
Clearly the $\Delta = 0.003$ case is viable
while the $\Delta = 0.006$ case is ruled out.

\section{Conclusions}

Precision electroweak measurements provide
stringent constraints on new physics beyond the SM.
For models which satisfy the three conditions listed
in section~2, the limits can be described in a model independent
way using the $STU$--formalism of Ref.~1.
Currently, the tightest limits come from the LEP/SLD observables,
with all other observables having little effect.
Current data also favors a smaller Higgs mass or a larger
$\alpha^{-1}(M_Z)$.
For models which are not encompassed within the $STU$ framework,
in particular, those with extra electroweak gauge bosons and/or
large vertex corrections, one can still obtain stringent
limits, albeit in a model dependent way, 
by focussing only on the vertex corrections at the $Z$ mass scale
and using the LEP/SLD observables to constrain them.

\section*{Acknowledgments}
We would like to thank M.~W.~Gr\"unewald for providing us
with the data used in this analysis.
This work was supported in part by the U. S. Department of Energy,
grant DE-FG05-92-ER40709, Task A (W.L.), and the
National Science Foundation, grant NSF~PHY-9802709 (A.K.G.).

\end{document}